\documentclass{article}
\usepackage{amsmath,amsthm,amssymb}
\newtheorem{theorem}{Theorem}[section]
\newtheorem{corollary}[theorem]{Corollary}
\newtheorem{lemma}[theorem]{Lemma}
\newtheorem{proposition}[theorem]{Proposition}
\theoremstyle{remark}
\newtheorem{remark}[theorem]{Remark}
\newtheorem{example}[theorem]{Example}
\numberwithin{equation}{section} \allowdisplaybreaks[3]
\newcommand\at{\allowbreak}

\title{Vlasov scaling for stochastic dynamics of~continuous~systems}

\author{Dmitri Finkelshtein\thanks{Institute of Mathematics,
 National Academy of Sciences of Ukraine, Kyiv, Ukraine ({\tt
 fdl@\at imath.\at kiev.\at ua}).}\and Yuri Kondratiev\thanks{Fakult\"{a}t
 f\"{u}r Mathematik, Universit\"{a}t Bielefeld, 33615 Bielefeld,
 Germany ({\tt kondrat@\at math.\at uni-bielefeld.\at de})}\and Oleksandr
 Kutoviy\thanks{Fakult\"{a}t f\"{u}r Mathematik, Universit\"{a}t
 Bielefeld, 33615 Bielefeld, Germany ({\tt
 kutoviy@\at math.\at uni-bielefeld.\at de}).}}

\newcommand\R{{\mathbb R}}
\newcommand\X{{\R^d}}

\newcommand\M{{\mathcal M}}
\newcommand\F{{\mathcal F}}
\newcommand\B{{\mathcal B}}
\newcommand\Bc{\B_{\mathrm{b}}}
\newcommand\Bbs{B_{\mathrm{bs}}}
\newcommand\Fc{{\F_{\mathrm{cyl}}}}
\renewcommand\L{{\mathcal L}}
\newcommand\K{{\mathcal K}}

\newcommand\La{\Lambda}
\newcommand\la{\lambda}
\newcommand\Ga{\Gamma}
\newcommand\ga{\gamma}
\newcommand\eps{\varepsilon}

\newcommand\n{{|\eta|}}
\newcommand\ren{{\eps, \mathrm{ren}}}

\newcommand\lu{\left\langle}
\newcommand\ru{\right\rangle}
\newcommand\llu{\lu\!\lu}
\newcommand\rru{\ru\!\ru}

\newcommand\h{\widehat}
\newcommand\jump{\mathrm{hop}}
\newcommand\bad{\mathrm{bad}}
\newcommand\tr{\triangle}
\newcommand\e{{(\eps)}}
\begin{document}

\maketitle

\begin{abstract}

We describe a general derivation scheme for the Vlasov-type
equations for Markov evolutions of particle systems in continuum.
This scheme is based on a proper scaling of corresponding Markov
generators and has an algorithmic realization in terms of  related
hierarchical chains of correlation functions equations. Several
examples of realization of the proposed approach in particular
models are presented.

\end{abstract}



\section{Introduction}

Dynamical processes in many-body systems are often approximately
described by kinetic equations, see, e.g., the excellent reviews by
H.Spohn \cite{Spo1980}, \cite{Spo1991}. A~famous example of such
equations is the Vlasov equation for a plasma, see e.g.
\cite{Neun1}, \cite{Neun2}. The~Vlasov equation in physics describes
the Hamiltonian motion of an infinite particle system in the
mean-field scaling limit, thereby taking into account the influence
of weak long-range forces. The convergence in the Vlasov scaling
limit was shown by W.Braun and K.Hepp \cite{BH1977} (for the
Hamiltonian dynamics) and by R.L.Dobrushin \cite{Dob1979} (for more
general deterministic dynamical systems). Note that the resulting
Vlasov-type equations for particle densities are considered in
classes of finite measures (in the weak form) or integrable
functions (in the strong form). The latter means, in fact, that we
are restricted to the case of finite-volume systems or systems with
zero mean density in an infinite volume. A detailed analysis of
Vlasov-type equations for integrable functions presented in the
recent paper by V.V.Kozlov \cite{Koz2008}.

The main aim of this paper is to study Vlasov-type scaling for some
classes of stochastic evolutions in continuum. Here we have in mind,
first of all,  spatial birth-and-death Markov processes (e.g.,
continuous Glauber dynamics) and hopping particles Markov evolutions
(e.g., Kawasaki dynamics in continuum).  Note that the approaches to
the Vlasov scaling mentioned above seems to be quite difficult to
apply to stochastic dynamics considered here (even in a finite
volume) due to some essential reasons. For these processes, the
possibility of their descriptions in terms of proper stochastic
evolutional equations for particle motion is generally speaking
absent. This, together with a possible variation of the particle
number during the evolution, is an essential trouble in the
application of the general Dobrushin's method.

Therefore, we shall look for an alternative approach to derive the
kinetic Vlasov-type equations from stochastic dynamics. Contrary to
the classical derivation of the Vlasov-type kinetic equations from
the Hamiltonian dynamics, we do not prove the law of large numbers
for the corresponding processes. We do not even need to show the
existence of the corresponding microscopic rescaled processes. Our
main idea is to study the evolution of states (distributions) of the
system in terms of the corresponding chain of hierarchical
equations. As pointed out by H.Spohn \cite{Spo1980}, the correct
Vlasov limit can be easily guessed from the BBGKY hierarchy for the
Hamiltonian system. Such heuristic derivation does not assume the
integrability condition for the density, but until now, it could not
be made rigorous due to the lack of detailed information about the
properties of solutions to the BBGKY hierarchy. We would like to
stress that different classes of initial data are not only
mathematical tools for the rigorous study of the problem. They
describe different physical situations in related microscopic
models. The zero average density systems were considered in
\cite{MaslovTereverdiev} by means of heuristic limit transition in
the corresponding hierarchical equations for correlation functions.
The framework we are working in is nonzero average density which is
related to the case of bounded correlation functions. Our approach
is based on Spohn's observation applied in a new dynamical
framework. More precisely, we already know that many  stochastic
evolutions in continuum admit effective descriptions in terms of
hierarchical equations for correlation functions which generalize
the BBGKY hierarchy from Hamiltonian to Markov setting, see, e.g.,
\cite{FKO2009} and the references therein. Moreover, these
hierarchical equations are often the only available technical tools
for the construction of corresponding dynamics in several models
\cite{KKM2008}, \cite{KKZ2006}, \cite{FKK2009}.

In Section 3 we propose a general scheme for the Vlasov scaling of
stochastic dynamics for interacting particle systems in continuum.
This scaling is actually of mean-field type which is adopted to
preserve the spatial structure. Additionally, we scale the class of
initial distributions at the level of the corresponding correlation
functions. The scheme  we use has also a clear interpretation in
terms of scaled Markov generators. An application of the considered
scaling leads to the limiting hierarchy which possesses a chaos
preservation property. Namely, if we start from a Poissonian
(non-homogeneous) initial state of the system, then this property
will be preserved during the time evolution. The main observation
which appears at this point is the following. A~special structure of
the interaction in the resulting virtual Vlasov system gives a
non-linear evolutional equation for the density of the evolving
Poisson state. It is for the first time that macroscopic Vlasov-type
equations are obtained from the microscopic infinite-particle
systems in an unbounded region of non-zero average density using the
corresponding system of hierarchical equations.

Section 4 is devoted to the application of the general scheme to
a~wide class of birth-and-death and hopping particles processes. We
state conditions on structural coefficients in the corresponding
Markov generators which give a~weak convergence of the rescaled
generators to the limiting generators of  the related Vlasov
hierarchies. As a result, we may  compute the limiting Vlasov-type
equations for the considered processes leaving the question about
the strong convergence of the hierarchy solutions open. In Section 5
we present a collection of particular examples of the resulting
Vlasov equations for several concrete models. Note that each of the
examples considered creates its own non-linear equation for the
density in the discussed scaling. These equations include
convolution operators as a common point of their structure. To our
knowledge, any general results concerning properties of solutions to
such kind of non-linear evolutional equation are absent. This is an
exiting mathematical problem strongly motivated by concrete models
of interacting particle dynamics.

Many problems of the (mathematical) population biology concerns
interactions between populations of different types. Our technic, in
fact, covers this case. In particular, one can derive spatially
inhomogeneous non-linear equations of the Lotka--Volterra type in
the Vlasov-type scaling. On the other hand, one may apply this
approach to the so-called continuous Ising model (Potts model). We
explain these results in forthcoming papers \cite{FKK2010d},
\cite{FKK2010e}.

Note that control of convergence of the Vlasov scalings for the
solutions to considered hierarchies is a difficult technical problem
which shall be analyzed for every particular model separately. Our
results in this direction concern two classes of models:  Glauber
dynamics in continuum and a spatial ecological model (so-called
Bolker--Dieckmann--Law--Pacala model). Due to their technically
complicated character, these results will be published in separated
works \cite{FKK2010b}, \cite{FKK2010c}.

\section{Basic facts and notation}

Let ${\mathcal{B}}({{\mathbb{R}}^d})$ be the family of all Borel sets in ${{%
\mathbb{R}}^d}$, $d\geq 1$; ${\mathcal{B}}_{\mathrm{b}}
({{\mathbb{R}}^d})$ denotes the system of all bounded sets in
${\mathcal{B}}({{\mathbb{R}}^d})$.

The configuration space over space ${{\mathbb{R}}^d}$ consists of
all locally finite subsets (configurations) of ${{\mathbb{R}}^d}$,
namely,
\begin{equation}  \label{confspace}
\Gamma =\Gamma_{{\mathbb{R}}^d} :=\bigl\{ \gamma \subset {{\mathbb{R}}^d} %
\bigm| |\gamma \cap \Lambda |<\infty, \text{ for all } \Lambda \in {\mathcal{%
B}}_{\mathrm{b}} ({{\mathbb{R}}^d})\bigr\}.
\end{equation}
The space $\Gamma$ is equipped with the vague topology, i.e., the
minimal topology for which all mappings $\Gamma\ni\gamma\mapsto
\sum_{x\in\gamma}
f(x)\in{\mathbb{R}}$ are continuous for any continuous function $f$ on ${{%
\mathbb{R}}^d}$ with compact support; note that the summation in
$\sum_{x\in\gamma} f(x)$ is taken over finitely many points of
$\gamma$ which belong to the support of $f$. In \cite{KK2006}, it
was shown that $\Gamma$ with the vague topology may be metrizable
and becomes a Polish space (i.e., complete separable metric space).
Corresponding to this topology, the Borel $\sigma $-algebra
${\mathcal{B}}(\Gamma )$ is the smallest $\sigma $-algebra
for which all mappings $\Gamma \ni \gamma \mapsto |\gamma_ \Lambda |\in{%
\mathbb{N}}_0:={\mathbb{N}}\cup\{0\}$ are measurable for any $\Lambda\in{%
\mathcal{B}}_{\mathrm{b}}({{\mathbb{R}}^d})$. Here $\gamma_\Lambda:=\gamma%
\cap\Lambda$, and $|\cdot|$ means the cardinality of a finite set.

The space of $n$-point configurations in an arbitrary
$Y\in{\mathcal{B}}({{\mathbb{R}}^d})$ is defined by
\begin{equation*}
\Gamma^{(n)}_Y:=\bigl\{  \eta \subset Y \bigm| |\eta |=n\bigr\} ,\quad n\in {%
\mathbb{N}}.
\end{equation*}
We set also $\Gamma^{(0)}_Y:=\{\emptyset\}$. As a set,
$\Gamma^{(n)}_Y$ may be identified with the symmetrization of
\begin{equation*}
\widetilde{Y^n} = \bigl\{ (x_1,\ldots ,x_n)\in Y^n \bigm| x_k\neq
x_l \text{ if } k\neq l\bigr\} .
\end{equation*}

Hence, one can introduce the corresponding Borel $\sigma $-algebra,
which we denote by ${\mathcal{B}}(\Gamma^{(n)}_Y)$. The space of
finite configurations in an arbitrary
$Y\in{\mathcal{B}}({{\mathbb{R}}^d})$ is defined by
\begin{equation*}
\Gamma_{0,Y}:=\bigsqcup_{n\in {\mathbb{N}}_0}\Gamma^{(n)}_Y.
\end{equation*}
This space is equipped with the topology of disjoint unions.
Therefore, one can introduce the corresponding Borel $\sigma
$-algebra ${\mathcal{B}} (\Gamma _{0,Y})$. In the case of
$Y={{\mathbb{R}}^d}$ we will omit the index $Y$ in
the notation, namely, $\Gamma_0:=\Gamma_{0,{{\mathbb{R}}^d}}$, $%
\Gamma^{(n)}:=\Gamma^{(n)}_{{{\mathbb{R}}^d}}$.

The restriction of the Lebesgue product measure $(dx)^n$ to $\bigl(%
\Gamma^{(n)}, {\mathcal{B}}(\Gamma^{(n)})\bigr)$ we denote by
$m^{(n)}$. We
set $m^{(0)}:=\delta_{\{\emptyset\}}$. The Lebesgue--Poisson measure $%
\lambda $ on $\Gamma_0$ is defined by
\begin{equation}  \label{LP-meas-def}
\lambda :=\sum_{n=0}^\infty \frac {1}{n!}m^{(n)}.
\end{equation}
For any $\Lambda\in{\mathcal{B}}_{\mathrm{b}}({{\mathbb{R}}^d})$ the
restriction of $\lambda$ to $\Gamma _\Lambda:=\Gamma_{0,\Lambda}$
will be also denoted by $\lambda $. The space $\bigl( \Gamma,
{\mathcal{B}}(\Gamma)\bigr)$ is the projective limit of the family
of spaces $\bigl\{( \Gamma_\Lambda, {\mathcal{B}}(\Gamma_\Lambda))\bigr\}%
_{\Lambda \in {\mathcal{B}}_{\mathrm{b}} ({{\mathbb{R}}^d})}$. The
Poisson measure $\pi$ on $\bigl(\Gamma ,{\mathcal{B}}(\Gamma
)\bigr)$ is given as the projective limit of the family of measures
$\{\pi^\Lambda \}_{\Lambda \in {\mathcal{B}}_{\mathrm{b}}
({{\mathbb{R}}^d})}$, where $\pi^\Lambda:=e^{-m(\Lambda)}\lambda $
is the probability measure on $\bigl( \Gamma_\Lambda,
{\mathcal{B}}(\Gamma_\Lambda)\bigr)$. Here $m(\Lambda)$ is the
Lebesgue measure of $\Lambda\in {\mathcal{B}}_{\mathrm{b}}
({{\mathbb{R}} ^d})$.

For any measurable function $f:{{\mathbb{R}}^d}\rightarrow
{\mathbb{R}}$ we define a \emph{Lebesgue--Poisson exponent}
\begin{equation}  \label{LP-exp-def}
e_\lambda(f,\eta):=\prod_{x\in\eta} f(x),\quad \eta\in\Gamma_0;
\qquad e_\lambda(f,\emptyset):=1.
\end{equation}
Then, by \eqref{LP-meas-def}, for $f\in L^1({{\mathbb{R}}^d},dx)$ we
obtain $e_\lambda(f)\in L^1(\Gamma_0,d\lambda)$ and
\begin{equation}  \label{LP-exp-mean}
\int_{\Gamma_0}
e_\lambda(f,\eta)d\lambda(\eta)=\exp\biggl\{\int_{{\mathbb{R}}^d}
f(x)dx\biggr\}.
\end{equation}

A set $M\in {\mathcal{B}} (\Gamma_0)$ is called bounded if there exists $%
\Lambda \in {\mathcal{B}}_{\mathrm{b}} ({{\mathbb{R}}^d})$ and $N\in {%
\mathbb{N}}$ such that $M\subset \bigsqcup_{n=0}^N\Gamma
_\Lambda^{(n)}$.
The set of bounded measurable functions with bounded support we denote by $%
B_{\mathrm{bs}}(\Gamma_0)$, i.e., $G\in B_{\mathrm{bs}}(\Gamma_0)$ if $%
G\upharpoonright_{\Gamma_0\setminus M}=0$ for some bounded $M\in {\mathcal{B}%
}(\Gamma_0)$. Any ${\mathcal{B}}(\Gamma_0)$-measurable function $G$ on $%
\Gamma_0$, in fact, is a sequence of functions $\bigl\{G^{(n)}\bigr\}_{n\in{%
\mathbb{N}}_0}$ where $G^{(n)}$ is a ${\mathcal{B}}(\Gamma^{(n)})$%
-measurable function on $\Gamma^{(n)}$. We consider also the set ${{\mathcal{%
F}}_{\mathrm{cyl}}}(\Gamma )$ of \textit{cylinder functions} on
$\Gamma$. Each $F\in {{\mathcal{F}}_{\mathrm{cyl}}}(\Gamma )$ is
characterized by the following relation: $F(\gamma
)=F\upharpoonright_{\Gamma_\Lambda
}(\gamma_\Lambda )$ for some $\Lambda\in {\mathcal{B}}_{\mathrm{b}}({{%
\mathbb{R}}^d})$.

There is the following mapping from $B_{\mathrm{bs}} (\Gamma_0)$ into ${{%
\mathcal{F}}_{\mathrm{cyl}}}(\Gamma )$, which plays the key role in
our further considerations:
\begin{equation}
KG(\gamma ):=\sum_{\eta \Subset \gamma }G(\eta ), \quad \gamma \in
\Gamma, \label{KT3.15}
\end{equation}
where $G\in B_{\mathrm{bs}}(\Gamma_0)$, see, e.g.,
\cite{KK2002,Len1975,Len1975a}. The summation in \eqref{KT3.15} is
taken over all finite subconfigurations $\eta\in\Ga_0$ of the
(infinite) configuration $\gamma\in\Ga$; we denote this by the
symbol, $\eta\Subset\gamma $. The mapping $K$ is linear, positivity
preserving, and invertible, with
\begin{equation}
K^{-1}F(\eta ):=\sum_{\xi \subset \eta }(-1)^{|\eta \setminus \xi
|}F(\xi ),\quad \eta \in \Gamma_0.  \label{k-1trans}
\end{equation}
Here and in the sequel inclusions like $\xi\subset\eta$ hold for $%
\xi=\emptyset$ as well as for $\xi=\eta$. We denote the restriction
of $K$ onto functions on $\Gamma_0$ by $K_0$.

For any fixed $C>1$ we consider the following Banach space of
${\mathcal{B}} (\Gamma_0)$-measurable functions
\begin{equation}  \label{norm}
\L _C:=\biggl\{ G:\Gamma_0\rightarrow{\mathbb{R}} \biggm| \|G\|_C:=
\int_{\Gamma_0} |G(\eta)| C^{|\eta|} d\lambda(\eta) <\infty\biggr\}.
\end{equation}

A measure $\mu \in {\mathcal{M}}_{\mathrm{fm} }^1(\Gamma )$ is
called locally absolutely continuous with respect to (w.r.t. for
short) the Poisson measure $\pi$ if for any $\Lambda \in
{\mathcal{B}}_{\mathrm{b}} ({{\mathbb{R}}^d})$ the projection of
$\mu$ onto $\Gamma_\Lambda$ is absolutely continuous w.r.t. the
projection of $\pi$ onto $\Gamma_\Lambda$. By \cite{KK2002}, in this
case, there exists a \emph{correlation functional} $k_{\mu}:\Gamma_0
\rightarrow {\mathbb{R}}_+$ such that for any $G\in B_{\mathrm{bs}}
(\Gamma_0)$ the following equality holds
\begin{equation}  \label{eqmeans}
\int_\Gamma (KG)(\gamma) d\mu(\gamma)=\int_{\Gamma_0}G(\eta)
k_\mu(\eta)d\lambda(\eta).
\end{equation}
The restrictions $k_\mu^{(n)}$ of this functional on $\Gamma_0^{(n)}$, $n\in{%
\mathbb{N}}_0$ are called \emph{correlation functions} of the
measure $\mu$. Note that $k_\mu^{(0)}=1$.

We recall now without a proof the special case of the well-known
technical lemma (cf., \cite{KMZ2004}) which plays very important
role in our calculations.

\begin{lemma}\label{Minlos} For any measurable function $H:\Gamma_0\times\Gamma_0\times%
\Gamma_0\rightarrow{\mathbb{R}}$
\begin{equation}  \label{minlosid}
\int_{\Gamma _{0}}\sum_{\xi \subset \eta }H\left( \xi ,\eta
\setminus \xi ,\eta \right) d\lambda \left( \eta \right)
=\int_{\Gamma _{0}}\int_{\Gamma _{0}}H\left( \xi ,\eta ,\eta \cup
\xi \right) d\lambda \left( \xi \right) d\lambda \left( \eta \right)
\end{equation}
only if both sides of the equality make sense.
\end{lemma}

\section{General scheme}

In this section we introduce the notion of the Vlasov scaling for
Markov dynamics of IPS on configuration spaces.

We assume that our system evolves in time due to some mechanism
whose details will be specified for  concrete models. Suppose that
the initial distribution of particles in our system is a measure
$\mu_0\in\M^1_\mathrm{fm}(\Ga)$, with correlation function $k_0$.
Let $\mu_t\in\M^1(\Ga)$ be the distribution of particles at time
$t>0$ and $k_t$ be its correlation function. One should note that if
evolution $(\mu_t)_{t\geq0}$ is ruled by an {\em \`a priori} given
Markov process on $\Ga$ (i.e. if such a Markov process exists), then
$\mu_t$ is a solution to the following Kolmogorov equation:
\[
\begin{cases}
\dfrac{d\mu_t}{dt}=L^\ast\mu_t\\
\mu_t\bigr\vert_{t=0}=\mu_0,
\end{cases}
\]
where $L^\ast$ is the operator adjoint to the generator of
functional evolution, i.e.,
\[
\begin{cases}
\dfrac{dF_t}{dt}=LF_t\\
F_t\bigr\vert_{t=0}=F_0.
\end{cases}
\]
Of course, one should be careful about the functional and measure
spaces to have all the above-introduced operators properly defined.
We postpone careful definitions of all these objects until the
introduction of concrete models.

Now, assume that the evolution of correlation functions
$(k_t)_{t\geq0}$, corresponding to $(\mu_t)_{t\geq0}$, first of all
exists, and is the solution of the following evolutional equation
\begin{equation}\label{corfeqn}
\begin{cases}
\dfrac{dk_t}{dt}=L^\tr k_t\\[3mm]
k_t\bigr\vert_{t=0}=k_0
\end{cases}
\end{equation}
where $L^\tr$ is the generator of a semigroup $T_t^\tr$ on some
functional space which includes all bounded functions (or bounded
with some weight) almost everywhere (a.e.) w.r.t. the
Lebesgue--Poisson measure $\la$. In many applications this space may
be taken to be $\K_C:=\bigl\{k:\Ga_0\rightarrow\R\bigm\vert k\cdot
C^{-\n} \in L^\infty (\la)\bigr\}$ for some  fixed $C>1$. Let us
stress that \eqref{corfeqn} is nothing else but a hierarchical
system of equations corresponding to the Markov generator
considered. This system has the same meaning as the BBGKY hierarchy
in the case of Hamiltonian dynamics.

The first important step on the way to construct the Vlasov scaling
concerns the proper rescaling of the initial state of the system.
Or, equivalently, in the language of correlation functions it means
the proper rescaling of the initial conditions of \eqref{corfeqn}.

More precisely, at the beginning we rescale $k_0$ with parameter
$\eps>0$ in such a way that the resulting functions $k_0^\e$ as
$\eps\rightarrow 0$ behave as follows:
\begin{equation}\label{scaling1}
k^\e_{0,\,\mathrm{ren}}(\eta):=\eps^\n k_0^\e(\eta)\rightarrow
r_0(\eta), \quad \eps\rightarrow0, \ \eta\in\Ga_0,
\end{equation}
where the function $r_0$ is a subject of choice for concrete
examples and aims. In general, it has to be a bounded function also
(or bounded with some weight) a.e. w.r.t. the Lebesgue--Poisson
measure.

\begin{remark}
In the case of $r_0(\eta)=e_\la(\rho_0,\eta)$, $\eta\in\Ga_0$,
$\rho_0:\X\rightarrow (0,+\infty)$ the assumption about the
rescaling of the initial condition means heuristically the
following: $\mu_{0,\,\mathrm{ren}}^\e\rightarrow \pi_{\rho_0}$,
where $\mu_{0,\,\mathrm{ren}}^\e$ has a correlation function
$\eps^\n k_0^\e(\eta)$.
\end{remark}

It is clear that such a rescaling of the initial solution for
\eqref{corfeqn} leads to a singular function w.r.t. $\eps>0$. In
applications, this fact can be interpreted as the growth of density
of the system with $\eps\rightarrow 0$.

We have to consider (and it is our second step) some proper scaling
of the generator in \eqref{corfeqn}:
\begin{equation}\label{operatorscaling}
L^\tr\longmapsto L^\tr_\eps.
\end{equation}
The concrete type of this scaling will depend on $L^\tr$. In the
next sections we consider several types of generators and
corresponding scalings. Suppose that there exists a solution of the
functional evolution
\begin{equation}\label{corfeqneps}
\begin{cases}
\dfrac{dk_t^\e}{dt}=L^\tr_\eps k_t^\e\\[3mm]
k_t^\e\bigr\vert_{t=0}=k_0^\e
\end{cases}
\end{equation}
We expect (and this will be shown in the concrete models for the
concrete scalings in forthcoming papers) that this solution will be
also singular w.r.t. $\eps>0$, hence, this solutions will be in
functional spaces depending on $\eps$.

Moreover, we should choose the type of scaling
\eqref{operatorscaling} which guarantees that the order of this
singularity will be the same for the initial function $k_0^\e$.
Namely (and it is our third step on the way to realize the Vlasov
scaling) we consider, cf. \eqref{scaling1},
\begin{equation}\label{scaling2}
k^\e_{t,\,\mathrm{ren}}(\eta):=\eps^\n k_t^\e(\eta), \quad
\eta\in\Ga_0,
\end{equation}
and we want to show that
\begin{equation}\label{limit2}
k^\e_{t,\,\mathrm{ren}}(\eta) \rightarrow r_t(\eta), \quad
\eps\rightarrow0, \ \eta\in\Ga_0.
\end{equation}

In fact, \eqref{scaling2} means that we consider a renormalized
version of the evolution equation \eqref{corfeqneps}:
\begin{equation}\label{corfeqnren}
\begin{cases}
\dfrac{dk_{t,\,\mathrm{ren}}^\e}{dt}=L^\tr_\ren k_{t,\,\mathrm{ren}}^\e\\[3mm]
k_{t,\,\mathrm{ren}}^\e\bigr\vert_{t=0}=k_{0,\,\mathrm{ren}}^\e
\end{cases}
\end{equation}
where
\begin{equation}\label{renorm_def}
L^\tr_\ren =\eps^\n L^\tr_\eps \eps^{-\n}.
\end{equation}

Therefore, informally, we want to show that the solution of the
evolution equation \eqref{corfeqnren} converges (in a proper sense)
to some function $r_t$ which satisfies the {\em Vlasov hierarchy}
\begin{equation}\label{corfeqnVl}
\begin{cases}
\dfrac{dr_t}{dt}=V^\tr r_t\\[3mm]
r_t\bigr\vert_{t=0}=r_0
\end{cases}
\end{equation}

Recall again that the choice of the scaling \eqref{operatorscaling}
is prescribed by the model. Having applications in mind, it is
important to consider the case of $r_0(\eta)=e_\la(\rho_0,\eta)$ and
the scaling \eqref{operatorscaling} which leads to $r_t$ of the same
type, i.e.,
\[
r_t(\eta)=e_\la(\rho_t,\eta), \quad \eta\in\Ga_0.
\]
The latter means the so-called chaos preservation property of the
Vlasov hierarchy. Equation \eqref{corfeqnVl} in this case implies,
in general, a non-linear equation for $\rho_t$:
\begin{equation}\label{V-eqn-gen}
\frac{\partial}{\partial t}\rho_t(x) = \upsilon(\rho_t)(x),\quad
x\in\X,
\end{equation}
which we will call the {\em Vlasov-type equation}.

To describe this scheme in a more analytical way, we use the
language of semigroups. Suppose that we know the mechanism of the
evolution of our system given by the Markov pre-generator $L$. Let
$L$ be defined at least on functions from $\Fc\left( \Gamma \right)
$ and
% $L^{\ast }$ be the dual mapping on
% $\mathcal{M}^{1}\left( \Gamma \right) $ w.r.t. duality $\left\langle
% F,\mu \right\rangle =\int_{\Gamma }F\left( \gamma \right) d\mu
% \left( \gamma \right) $. Let
$\h{L}=K^{-1}LK$ be the corresponding descend mapping on functions
from $\Bbs\left( \Gamma _{0}\right)$. Let us fix the duality between
functions on $\Ga_0$
\begin{equation}\label{duality}
\llu G,k\rru =\int_{\Gamma _{0}}G\left( \eta \right) k\left( \eta
\right) d\lambda \left( \eta \right),
\end{equation}
and consider the mapping $L^\tr$ being the dual to $\h{L}$ w.r.t.
\eqref{duality}.

Assume that $L$ can be extended to a generator $L$. We want to
construct a scaling of the generator $L$, say, $L_\eps$, $\eps>0$,
such that the scheme described above will be covered. Assume that we
have a semigroup $\h{T}_\eps(t)$ with a generator
$\h{L}_\eps=K^{-1}L_\eps K$ in some functional space over $\Ga_0$.
Consider the dual semigroup $T^\tr_\eps(t)$ which corresponds (in a
proper sense) to $L^\tr_\eps$. As we said before, we consider an
initial condition of \eqref{corfeqneps} with a singularity in
$\eps$, namely, $k_0^{(\eps)}(\eta) \sim \eps^{-|\eta|} r_0(\eta)$,
$\eps\rightarrow 0$, $\eta\in\Ga_0$ with some function $r_0$,
independent of $\eps$. First of all, we have to choose such a
scaling $L\mapsto L_\eps$ for which $T^\tr_\eps(t)$ preserves the
order of the singularity:
\begin{equation}\label{dream}
(T^\tr_\eps(t)k_0^{(\eps)})(\eta) \sim \eps^{-|\eta|} r_t(\eta),
\quad \eps\rightarrow 0, \ \eta\in\Ga_0.
\end{equation}
And the most important is that the dynamics $r_0 \mapsto r_t$ should
preserve the Lebesgue--Poisson exponents: if
$r_0(\eta)=e_\la(\rho_0,\eta)$ then $r_t(\eta)=e_\la(\rho_t,\eta)$,
where $\rho_t$ is satisfied \eqref {V-eqn-gen}.

Now let us close our construction with the evolution of states in
this scheme. Let us consider for any $\eps>0$ the following mapping
of functions on $\Ga_0$
\begin{equation}
(R_\eps r)(\eta):=\eps^\n r(\eta).
\end{equation}
This mapping is ``self-dual'' w.r.t. duality \eqref{duality},
moreover, $R_\eps^{-1}=R_{\eps^{-1}}$. Then we have
$k^{(\eps)}_0\sim R_{\eps^{-1}} r_0$, and we need $r_t \sim R_\eps
T^\tr_\eps(t)k_0^{(\eps)} \sim  R_\eps T^\tr_\eps(t)R_{\eps^{-1}}
r_0$. Therefore, we have to show that for any $t\geq 0$ the operator
family $R_\eps T^\tr_\eps(t)R_{\eps^{-1}}$, $\eps>0$ has a limiting
(in a proper sense) operator $U(t)$ as $\eps\rightarrow 0$ and
\begin{equation}\label{chaospreserving}
U(t)e_\la(\rho_0)=e_\la(\rho_t).
\end{equation}
But, informally, $T^\tr_\eps(t)=\exp{\{tL^\tr_\eps\}}$ and $R_\eps
T^\tr_\eps(t)R_{\eps^{-1}}=\exp{\{t R_\eps L^\tr_\eps R_{\eps^{-1}}
\}}$. In fact, we need the existence of an operator $V^\tr$ such
that $\exp{\{t R_\eps L^\tr_\eps R_{\eps^{-1}} \}}\rightarrow
\exp{\{tV^\tr\}=:U(t)}$ for which \eqref{chaospreserving} holds.
Therefore, a heuristic way to produce the scaling $L\mapsto L_\eps$
is to demand that
\begin{equation}\label{demand}
\lim_{\eps\rightarrow 0}\left(\frac{\partial}{\partial
t}e_\la(\rho_t,\eta)-L^\tr_\ren e_\la(\rho_t,\eta)\right)=0, \quad
\eta\in\Ga_0
%=\sum_{x\in\eta}\frac{\partial}{\partial t}
%\rho_t(x) e_\la(\rho_t,\eta\setminus x)-L^\tr_\ren
%e_\la(\rho_t,\eta)
\end{equation}
if $\rho_t$ is satisfied \eqref{V-eqn-gen}. The point-wise limit of
$L^\tr_\ren$ will be the natural candidate for $V^\tr$.

Sometimes, to show convergence of  solutions of evolutional
equations in some functional spaces it is much simpler to work with
the operators $\h{L}_\ren$ and $\h{V}$ which are pre-dual to
$L^\tr_\ren$ and $V^\tr$ w.r.t. the duality \eqref{duality}. Note
that \eqref{renorm_def} implies
\begin{equation}\label{ren_desc}
\h{L}_\ren=R_{\eps^{-1}}\h{L}_\eps R_\eps,
\end{equation}
and $\h{V}$ should be the point-wise limit of $\h{L}_\ren$.

\section{Generators of birth, death, and hopping}

Through out this section we consider generators of two types for
continuous models: the birth-and-death generator $L_{\bad}= L^- +
L^+$ and the hopping generator $L_{\jump}$, where for any
$F\in\Fc(\Ga)$
\begin{align}
\bigl( L^{-}F\bigr) \left( \gamma \right)  :=&\sum_{x\in \gamma
}d\left( x,\gamma \setminus x\right) \left[ F\left( \gamma \setminus
x\right)
-F\left( \gamma \right) \right],  \label{death-gen}\\
\bigl( L^{+}F\bigr) \left( \gamma \right)
:=&\int_{\mathbb{R}^{d}}b\left( x,\gamma \right) \left[ F\left(
\gamma \cup x\right) -F\left( \gamma \right) \right] dx,\label{birth-gen}\\
\left( L_{\jump}F\right) \left( \gamma \right) :=&\sum_{x\in \gamma }\int_{\mathbb{R}%
^{d}}c\left( x,y,\gamma \right) \left[ F\left( \gamma \setminus
x\cup y\right) -F\left( \gamma \right) \right] dy.\label{hop-gen}
\end{align}%
Here $b,d,c$ are measurable functions of their variables and,
additionally, $b$ and $c$ are locally integrable function of the
first and second variables, correspondingly. These conditions
guarantee that \eqref{death-gen}--\eqref{hop-gen} are well-defined
on $\Fc(\Ga)$ since for any $F\in\Fc(\Ga)$ there exists some
$\La\in\Bc(\X)$ such that $F(\ga\setminus x)=F(\ga)$ for any
$x\in\ga_{\La^c}$, $F(\ga\cup x)=F(\ga)$ for any $x\in\La^c$, and
$F(\ga\setminus x\cup y)=F(\ga)$ for any $x\in\ga_{\La^c}$,
$y\in\La^c$; as result the sums in \eqref{death-gen} and
\eqref{hop-gen} are over finite set $\ga_{\La}$ and the integrals in
\eqref{birth-gen} and \eqref{hop-gen} are over bounded set $\La$.

We may denote $L^-=L^-(d)$, $L^+=L^+(b)$, $L_{\jump}=L_{\jump}(c)$.
Assume that we have some scaling of rates $b,d,c$, say, $b_\eps,
d_\eps, c_\eps$, correspondingly; $\eps>0$. Then, let us consider
the following scaling of $L_\bad$ and $L_\jump$:
\begin{align}
L_{\bad,\, \eps}&=L^-(d_\eps)+\eps^{-1} L^+(b_\eps),
\label{bad-scaling}\\
L_{\jump,\, \eps} &= L_\jump(c_\eps). \label{hop-scaling}
\end{align}

\begin{remark}
In a conservative system with a generator like \eqref{hop-gen} which
preserves the ``number of particles'' during an evolution the
Vlasov-type scaling usually means decreasing of the intensity of the
interactions between elements of a system together with increasing
of correlations in the initial state. However, in a non-conservative
birth-and-death dynamics with a generator $L_{\bad}$ we need an
additional increasing of the birth intensity to preserve the
influence of the birth part in the limiting Vlasov hierarchy. Note
that the necessity of the concrete factor $\eps^{-1}$ in
\eqref{bad-scaling} is clear {\em a~posteriori} only (see
Proposition~\ref{undercond2}).
\end{remark}

Suppose that there exists three families of measurable functions on
$\Ga_0$: $D_{x}^{(\eps)}$, $B_{x}^{(\eps)}$, $C_{x,y}^{(\eps)}$,
$\eps>0$, $\{x,y\}\subset\X$, such that
\begin{gather*}
d_\eps\left( x,\ga \right) =(KD_{x}^{(\eps)})(\ga),\qquad
b_\eps\left( x,\ga \right) =(KB_{x}^{(\eps)})(\ga), \\ c_\eps\left(
x,y, \ga\right) = (K C_{x,y}^{(\eps)})(\ga\setminus x).
\end{gather*}
Note that, in general, $C_{x,y}\neq C_{y,x}$.

\begin{proposition}\label{renormgen}
The following formulas hold for any $k\in \Bbs(\Ga_0)$
\begin{align}
\bigl( L^\tr_{\bad,\ren} k\bigr) \left( \eta \right) =&-\int_{\Gamma
_{0}}k\left( \xi \cup \eta \right) \sum_{x\in \eta }\sum_{\omega
\subset \eta \setminus x}\varepsilon ^{-\left\vert \xi \right\vert
}D_{x}^{\left( \varepsilon \right) }\left( \omega \cup \xi
\right) d\lambda \left( \xi \right)  \label{bad-renorm}\\
&+\int_{\Gamma _{0}}\sum_{x\in \eta }k\left( \xi \cup \left( \eta
\setminus x\right) \right) \sum_{\omega \subset \eta \setminus
x}\varepsilon ^{-\left\vert \xi \right\vert }B_{x}^{\left(
\varepsilon \right) }\left( \omega \cup \xi \right) d\lambda \left(
\xi \right) ;\notag \\
\bigl( L^\tr_{\jump,\ren} k\bigr) \left( \eta \right) =&\sum_{x\in
\eta }\int_{\mathbb{R}^{d}}\int_{\Gamma _{0}}k\left( \xi \cup (\eta
\setminus x)\cup y\right) \sum_{\omega \subset \eta \setminus
x}\varepsilon ^{-\left\vert \xi \right\vert }C_{y,x}^{\left(
\varepsilon
\right) }\left( \omega \cup \xi \right) d\lambda \left( \xi \right) dy \notag\\
&-\sum_{x\in \eta }\int_{\Gamma _{0}}k\left( \xi \cup \eta \right)
\int_{\mathbb{R}^{d}}\sum_{\omega \subset \eta \setminus
x}\varepsilon ^{-\left\vert \xi \right\vert }C_{x,y}^{\left(
\varepsilon \right) }\left( \omega \cup \xi \right) dyd\lambda
\left( \xi \right) .\label{hop-renorm}
\end{align}
\end{proposition}
\begin{proof}
The proof is straightforward. By \cite{FKO2009}, from
\eqref{bad-scaling} and \eqref{hop-scaling} we have
\begin{align*}
\bigl( L^\tr_{\bad,\eps} k\bigr)\left( \eta \right)=&-\int_{\Gamma
_{0}}k\left( \xi \cup \eta \right) \sum_{x\in \eta }\sum_{\omega
\subset \eta \setminus x}D_{x}^{\left( \varepsilon \right)
}\left( \omega \cup \xi \right) d\lambda \left( \xi \right)  \\
&+\varepsilon ^{-1}\int_{\Gamma _{0}}\sum_{x\in \eta }k\left( \xi
\cup \left( \eta \setminus x\right) \right) \sum_{\omega \subset
\eta \setminus x}B_{x}^{\left( \varepsilon \right) }\left( \omega
\cup \xi \right) d\lambda \left( \xi \right);\\
\bigl( L^\tr_{\jump,\eps} k\bigr)  \left( \eta \right)  =&\sum_{y\in \eta }\int_{%
\mathbb{R}^{d}}\int_{\Gamma _{0}}k\left( \xi \cup (\eta \setminus
y)\cup x\right) \sum_{\omega \subset \eta \setminus
y}C^{(\eps)}_{x,y}\left( \omega \cup \xi
\right) d\lambda \left( \xi \right) dx \\
&-\int_{\Gamma _{0}}k\left( \xi \cup \eta \right) \sum_{y\in \eta
}\sum_{\omega \subset \eta \setminus
y}\int_{\mathbb{R}^{d}}C_{y,x}^{(\eps)}\left( \omega \cup \xi
\right) dxd\lambda \left( \xi \right).
\end{align*}
Then, \eqref{bad-renorm} and \eqref{hop-renorm} follow
directly from \eqref{renorm_def}. %\qed
\end{proof}

Let $\rho_t$, $t\geq0$ be measurable functions on $\X$. The explicit
formula
\begin{equation}\label{der_exp}
\frac{\partial }{\partial t}e_{\lambda }\left( \rho _{t},\eta
\right)
=\sum_{x\in \eta }e_{\lambda }\left( \rho _{t},\eta \setminus x\right) \frac{%
\partial }{\partial t}\rho _{t}\left( x\right)
\end{equation}
together with our ``demand'' \eqref{demand} induce us to state the
following corollary.
\begin{corollary}
Let $\rho$ be a measurable function on $\X$. Then
\begin{align}
&\bigl( L^\tr_{\bad,\ren} e_\la(\rho)\bigr) \left( \eta
\right) \label{bad_ren_exp}\\
=&-\sum_{x\in \eta }e_{\lambda }\left( \rho ,\eta \setminus x\right)
\rho \left( x\right) \int_{\Gamma _{0}}e_{\lambda }\left( \rho ,\xi
\right) \sum_{\omega \subset \eta \setminus x}\varepsilon
^{-\left\vert \xi \right\vert }D_{x}^{\left( \varepsilon \right)
}\left( \omega \cup \xi
\right) d\lambda \left( \xi \right)  \notag\\
&+\sum_{x\in \eta }e_{\lambda }\left( \rho ,\eta \setminus x\right)
\int_{\Gamma _{0}}e_{\lambda }\left( \rho ,\xi \right) \sum_{\omega
\subset \eta \setminus x}\varepsilon ^{-\left\vert \xi \right\vert
}B_{x}^{\left( \varepsilon \right) }\left( \omega \cup \xi \right)
d\lambda \left( \xi \right)  ; \notag
\end{align}
and
\begin{align}
&\bigl( L^\tr_{\jump,\ren} e_\la(\rho) \bigr) \left( \eta
\right) \label{hop_ren_exp}\\=&\sum_{x\in \eta }e_{\lambda }\left( \rho ,\eta \setminus x\right) \int_{%
\mathbb{R}^{d}}\rho \left( y\right) \int_{\Gamma _{0}}e_{\lambda
}\left( \rho ,\xi \right) \sum_{\omega \subset \eta \setminus
x}\varepsilon ^{-\left\vert \xi \right\vert }C_{y,x}^{\left(
\varepsilon \right) }\left(
\omega \cup \xi \right) d\lambda \left( \xi \right) dy \notag\\
&-\sum_{x\in \eta }e_{\lambda }\left( \rho ,\eta \setminus x\right)
\rho \left( x\right) \int_{\mathbb{R}^{d}}\int_{\Gamma
_{0}}e_{\lambda }\left( \rho ,\xi \right) \sum_{\omega \subset \eta
\setminus x}\varepsilon ^{-\left\vert \xi \right\vert
}C_{x,y}^{\left( \varepsilon \right) }\left( \omega \cup \xi \right)
d\lambda \left( \xi \right) dy.\notag
\end{align}
\end{corollary}
\begin{proposition}\label{undercond}
Suppose that for any $\{x,y\}\subset\X$, $\{\xi,\eta\}\subset\Ga_0$
\begin{align}
\exists\,\lim_{\varepsilon \rightarrow 0}\sum_{\omega \subset \eta
}\varepsilon ^{-\left\vert \xi \right\vert }D_{x}^{\left(
\varepsilon \right) }\left( \omega \cup \xi \right)
=\lim_{\varepsilon \rightarrow 0}\varepsilon ^{-\left\vert \xi
\right\vert }D_{x}^{\left( \varepsilon \right) }\left( \xi
\right) =:&D_{x}^{V}(\xi), \label{dVl}\\
\exists\,\lim_{\varepsilon \rightarrow 0}\sum_{\omega \subset \eta
}\varepsilon ^{-\left\vert \xi \right\vert }B_{x}^{\left(
\varepsilon \right) }\left( \omega \cup \xi \right)
=\lim_{\varepsilon \rightarrow 0}\varepsilon ^{-\left\vert \xi
\right\vert }B_{x}^{\left( \varepsilon \right) }\left( \xi \right)
=:&B_{x}^{V}(\xi), \label{bVl}\\\exists\,\lim_{\varepsilon
\rightarrow 0}\sum_{\omega \subset \eta }\varepsilon ^{-\left\vert
\xi \right\vert }C_{x,y}^{\left( \varepsilon \right) }\left( \omega
\cup \xi \right) =\lim_{\varepsilon \rightarrow 0}\varepsilon
^{-\left\vert \xi \right\vert }C_{x,y}^{\left( \varepsilon \right)
}\left( \xi \right) =:&C_{x,y}^{V}(\xi). \label{cVl}
\end{align}
Then, our ``demand'' \eqref{demand} holds. More precisely,
\begin{align}\label{bad-Vlgen}
( V^\tr_{\bad} k) \left( \eta \right):=&\lim_{\eps\rightarrow
0}\bigl( L^\tr_{\bad,\ren} k\bigr) \left( \eta
\right)\\
=&-\int_{\Gamma _{0}}k\left( \xi \cup \eta \right) \sum_{x\in \eta
}D_{x}^{V}(\xi) d\lambda \left( \xi \right)  \notag\\
&+\int_{\Gamma _{0}}\sum_{x\in \eta }k\left( \xi \cup \left( \eta
\setminus x\right) \right)B_{x}^{V}(\xi)d\lambda \left( \xi \right)
,\notag
\end{align}
and if $\rho_t$ is the solution of the equation \eqref{V-eqn-gen}
with
\begin{align}
\upsilon(\rho)(x)=\upsilon_\bad(\rho)(x)=&-\rho \left( x\right)
\int_{\Gamma _{0}}e_{\lambda }\left( \rho ,\xi \right)
D_{x}^{V}\left( \xi \right) d\lambda \left( \xi \right)  \notag\\&+
\int_{\Gamma _{0}}e_{\lambda }\left( \rho ,\xi \right) B_{x}^{V
}\left(  \xi \right) d\lambda \left( \xi \right) ,
\end{align}
then, $ \dfrac{\partial}{\partial t}e_\la(\rho_t,\eta)=\bigl(
V^\tr_{\bad} e_{\la}(\rho_t)\bigr) \left( \eta \right)$.
Analogously,
\begin{align}
\label{hop-Vlgen} ( V^\tr_{\jump} k)  \left( \eta
\right):=&\lim_{\eps\rightarrow 0}\bigl( L^\tr_{\jump,\ren} k\bigr)
\left( \eta
\right)\\
=&\sum_{x\in \eta }\int_{\mathbb{R}^{d}}\int_{\Gamma _{0}}k\left(
\xi \cup(\eta \setminus x)\cup y\right) C_{y,x}^{V}\left( \xi \right) d\lambda \left( \xi \right) dy\notag \\
&-\sum_{x\in \eta }\int_{\Gamma _{0}}k\left( \xi \cup \eta \right)
\int_{\mathbb{R}^{d}}C_{x,y}^{V}\left( \xi \right) dyd\lambda \left(
\xi \right),\notag
\end{align}
and if $\rho_t$ is the solution of the equation \eqref{V-eqn-gen}
with
\begin{align}
\upsilon(\rho)(x)=\upsilon_\jump(\rho)(x)=&\int_{%
\mathbb{R}^{d}}\rho \left( y\right) \int_{\Gamma _{0}}e_{\lambda
}\left( \rho ,\xi \right) C_{y,x}^{V}\left( \xi \right)\ d\lambda
\left( \xi \right) dy \\ &- \rho \left( x\right)\int_{\Gamma
_{0}}e_{\lambda }\left( \rho ,\xi \right)
\int_{\mathbb{R}^{d}}C_{x,y}^{V}\left( \xi \right)dy d\lambda \left(
\xi \right) ,\notag
\end{align}
then, $ \dfrac{\partial}{\partial t}e_\la(\rho_t,\eta)=\bigl(
V^\tr_{\jump} e_{\la}(\rho_t)\bigr) \left( \eta \right)$.
\end{proposition}
\begin{proof}
The equalities \eqref{bad-Vlgen} and \eqref{hop-Vlgen} are direct
consequences of the Proposition~\ref{renormgen} and the conditions
\eqref{dVl}--\eqref{cVl}. Taking the limit in \eqref{bad_ren_exp}
and \eqref{hop_ren_exp} as $\eps\rightarrow 0$ and using
\eqref{der_exp} we obtain the statement. %\qed
\end{proof}

And now we present the explicit expressions for the corresponding
operators $\h{L}_\ren$ and $\h{V}$.

\begin{proposition}\label{undercond2}
For any $G\in\Bbs(\Ga_0)$ the following formulas hold
\begin{align}
\bigl( \h{L}_{\bad,\,\ren}G\bigl) \left( \eta \right) =&-\sum_{x\in
\eta }\sum_{\xi \subset \eta \setminus x}G\left( \xi \cup x\right)
\sum_{\omega \subset \xi }\varepsilon ^{-\left\vert (\eta\setminus
x)\setminus \xi \right\vert }D_{x}^{\left( \varepsilon \right)
}\left( \omega \cup (\eta \setminus x)\setminus \xi \right)
\notag\\\label{bad-desc-ren} &+\sum_{\xi \subset \eta
}\int_{\mathbb{R}^{d}}G\left( \xi \cup x\right) \sum_{\omega \subset
\xi }\varepsilon ^{-\left\vert \eta \setminus \xi \right\vert
}B_{x}^{\left( \varepsilon \right) }\left( \omega \cup \eta \setminus \xi \right) dx;\\
\bigl( \h{L}_{\jump,\,\ren}G\bigr) \left( \eta \right) =&\sum_{y\in
\eta }\sum_{\xi \subset \eta \setminus
y}\int_{\mathbb{R}^{d}}G\left( \xi \cup x\right) \sum_{\omega
\subset \xi }\varepsilon ^{-\left\vert (\eta \setminus y)\setminus
\xi \right\vert }C_{y,x}^{\left( \varepsilon \right) }\left( \omega
\cup (\eta \setminus y)\setminus \xi
\right) dx \notag \\
& -\sum_{\xi \subset \eta }G\left( \xi \right) \sum_{x\in \xi
}\sum_{\omega \subset \xi \setminus
x}\int_{\mathbb{R}^{d}}\varepsilon ^{-\left\vert \eta \setminus \xi
\right\vert }C_{x,y}^{\left( \varepsilon \right) }\left( \omega \cup
\eta \setminus \xi \right) dy. \label{jump-desc-ren}
\end{align}
If, additionally, \eqref{dVl}--\eqref{cVl} hold, then,
\begin{align}
\bigl( \h{V}_\bad G\bigr) \left( \eta \right) =&-\sum_{\xi \subset
\eta }G\left( \xi \right) \sum_{x\in \xi }D_{x}^{V}(\eta \setminus
\xi )\notag\\& +\sum_{\xi \subset \eta }\int_{\mathbb{R}^{d}}G\left(
\xi \cup
x\right) B_{x}^{V}(\eta \setminus \xi )dx;\label{bad-desc-Vl}\\
\bigl( \h{V}_\jump G\bigr) \left( \eta \right) =&\sum_{y\in \eta
}\sum_{\xi \subset \eta \setminus y}\int_{\mathbb{R}^{d}}G\left( \xi
\cup x\right) C_{y,x}^{V}\left( (\eta \setminus y)\setminus \xi
\right) dx \notag \\
& -\sum_{\xi \subset \eta }G\left( \xi \right) \sum_{x\in \xi
}\int_{\mathbb{R}^{d}}C_{x,y}^{V }\left( \eta \setminus \xi \right)
dy.\label{jump-desc-Vl}
\end{align}
\end{proposition}
\begin{proof}
We may obtain these formulas directly from the duality
\eqref{duality} and the~Lemma~\ref{Minlos}. Namely, for any
$G\in\Bbs(\Ga_0)$ we have
\begin{align*}
&\int_{\Gamma _{0}}G\left( \eta \right) \bigl( \h{L}_{\bad,\,\ren}k\bigr) \left( \eta \right) d\lambda \left( \eta \right) \\
=&-\int_{\Gamma _{0}}G\left( \eta \right) \int_{\Gamma _{0}}k\left(
\xi \cup \eta \right) \sum_{x\in \eta }\sum_{\omega \subset \eta
\setminus x}\varepsilon ^{-\left\vert \xi \right\vert }D_{x}^{\left(
\varepsilon \right) }\left( \omega \cup \xi \right) d\lambda \left(
\xi \right) d\lambda
\left( \eta \right) \\
&+\int_{\Gamma _{0}}G\left( \eta \right) \int_{\Gamma
_{0}}\sum_{x\in \eta }k\left( \xi \cup \left( \eta \setminus
x\right) \right) \sum_{\omega \subset \eta \setminus x}\varepsilon
^{-\left\vert \xi \right\vert }B_{x}^{\left( \varepsilon \right)
}\left( \omega \cup \xi \right) d\lambda
\left( \xi \right) d\lambda \left( \eta \right) \\
=&-\int_{\Gamma _{0}}\int_{\Gamma _{0}}\int_{\mathbb{R}^{d}}G\left(
\eta \cup x\right) k\left( \xi \cup \eta \cup x\right) \sum_{\omega
\subset \eta }\varepsilon ^{-\left\vert \xi \right\vert
}D_{x}^{\left( \varepsilon \right) }\left( \omega \cup \xi \right)
dxd\lambda \left( \xi \right)
d\lambda \left( \eta \right) \\
&+\int_{\Gamma _{0}}\int_{\Gamma _{0}}\int_{\mathbb{R}^{d}}G\left(
\eta \cup x\right) k\left( \xi \cup \eta \right) \sum_{\omega
\subset \eta }\varepsilon ^{-\left\vert \xi \right\vert
}B_{x}^{\left( \varepsilon \right) }\left( \omega \cup \xi \right)
dxd\lambda \left( \xi \right)
d\lambda \left( \eta \right) \\
=&-\int_{\Gamma _{0}}\int_{\mathbb{R}^{d}}\sum_{\eta \subset \xi
}G\left( \eta \cup x\right) k\left( \xi \cup x\right) \sum_{\omega
\subset \eta }\varepsilon ^{-\left\vert \xi \setminus \eta
\right\vert }D_{x}^{\left( \varepsilon \right) }\left( \omega \cup
\xi \setminus \eta \right)
dxd\lambda \left( \xi \right) \\
&+\int_{\Gamma _{0}}\sum_{\eta \subset \xi
}\int_{\mathbb{R}^{d}}G\left( \eta \cup x\right) k\left( \xi \right)
\sum_{\omega \subset \eta }\varepsilon ^{-\left\vert \xi \setminus
\eta \right\vert }B_{x}^{\left( \varepsilon \right) }\left( \omega
\cup \xi \setminus \eta \right)
dxd\lambda \left( \xi \right) \\
=&-\int_{\Gamma _{0}}\sum_{x\in \xi }\sum_{\eta \subset \xi
\setminus x}G\left( \eta \cup x\right) k\left( \xi \right)
\sum_{\omega \subset \eta }\varepsilon ^{-\left\vert \xi \setminus
x\setminus \eta \right\vert }D_{x}^{\left( \varepsilon \right)
}\left( \omega \cup \xi \setminus
x\setminus \eta \right) d\lambda \left( \xi \right) \\
&+\int_{\Gamma _{0}}\sum_{\eta \subset \xi
}\int_{\mathbb{R}^{d}}G\left( \eta \cup x\right) k\left( \xi \right)
\sum_{\omega \subset \eta }\varepsilon ^{-\left\vert \xi \setminus
\eta \right\vert }B_{x}^{\left( \varepsilon \right) }\left( \omega
\cup \xi \setminus \eta \right) dxd\lambda \left( \xi \right),
\end{align*}
which implies \eqref{bad-desc-ren}. To get \eqref{bad-desc-Vl} we
may proceed in the same way or just let $\eps\rightarrow 0$ in
\eqref{bad-desc-ren}. Then \eqref{dVl}--\eqref{bVl} together with
equality
\[
\sum_{x\in \eta }\sum_{\xi \subset \eta\setminus x }G\left( \xi\cup
x \right) D_{x}^{V}((\eta\setminus x) \setminus \xi ) =\sum_{\xi
\subset \eta }G\left( \xi \right) \sum_{x\in \xi }D_{x}^{V}(\eta
\setminus \xi )
\]
provide \eqref{bad-desc-Vl}.

Analogously, for any $G\in\Bbs(\Ga_0)$ we have
\begin{align*}
&\int_{\Gamma _{0}}G\left( \eta \right) \left( L_{\jump,\ren}^{\ast }k\right) \left( \eta \right) d\lambda \left( \eta \right)  \\
=& \int_{\Gamma _{0}}\int_{\mathbb{R}^{d}}G\left( \eta \cup x\right)
\int_{\mathbb{R}^{d}}\int_{\Gamma _{0}}k\left( \xi \cup \eta \cup
y\right) \sum_{\omega \subset \eta }\varepsilon ^{-\left\vert \xi
\right\vert }C_{y,x}^{\left( \varepsilon \right) }\left( \omega \cup
\xi \right)
d\lambda \left( \xi \right) dxdyd\lambda \left( \eta \right)  \\
& -\int_{\Gamma _{0}}G\left( \eta \right) \int_{\Gamma _{0}}k\left(
\xi \cup
\eta \right) \sum_{x\in \eta }\sum_{\omega \subset \eta \setminus x}\int_{%
\mathbb{R}^{d}}\varepsilon ^{-\left\vert \xi \right\vert
}C_{x,y}^{\left( \varepsilon \right) }\left( \omega \cup \xi \right)
dyd\lambda \left( \xi
\right) d\lambda \left( \eta \right)  \\
 =&\int_{\Gamma _{0}}\int_{\mathbb{R}^{d}}\int_{\mathbb{R}^{d}}\sum_{\eta
\subset \xi }G\left( \eta \cup x\right) k\left( \xi \cup y\right)
\sum_{\omega \subset \eta }\varepsilon ^{-\left\vert \xi \setminus
\eta \right\vert }C_{y,x}^{\left( \varepsilon \right) }\left( \omega
\cup \xi
\setminus \eta \right) dxdyd\lambda \left( \xi  \right)  \\
& -\int_{\Gamma _{0}}\sum_{\eta \subset \xi }G\left( \eta \right)
k\left(
\xi \right) \sum_{x\in \eta }\sum_{\omega \subset \eta \setminus x}\int_{%
\mathbb{R}^{d}}\varepsilon ^{-\left\vert \xi \setminus \eta
\right\vert }C_{x,y}^{\left( \varepsilon \right) }\left( \omega \cup
\xi \setminus \eta
\right) dyd\lambda \left( \xi  \right)  \\
=& \int_{\Gamma _{0}}k\left( \xi \right)
\int_{\mathbb{R}^{d}}\sum_{y\in \xi }\sum_{\eta \subset \xi
\setminus y}G\left( \eta \cup x\right) \sum_{\omega \subset \eta
}\varepsilon ^{-\left\vert \xi \setminus y\setminus \eta \right\vert
}C_{y,x}^{\left( \varepsilon \right) }\left( \omega \cup \xi
\setminus y\setminus \eta \right) dxd\lambda \left( \xi  \right)  \\
& -\int_{\Gamma _{0}}k\left( \xi \right) \sum_{\eta \subset \xi
}G\left(
\eta \right) \sum_{x\in \eta }\sum_{\omega \subset \eta \setminus x}\int_{%
\mathbb{R}^{d}}\varepsilon ^{-\left\vert \xi \setminus \eta
\right\vert }C_{x,y}^{\left( \varepsilon \right) }\left( \omega \cup
\xi \setminus \eta \right) dyd\lambda \left( \xi \right),
\end{align*}
which implies \eqref{jump-desc-ren}. To get \eqref{jump-desc-Vl} we
may proceed again in the same way or just let $\eps\rightarrow 0$ in
\eqref{jump-desc-ren} and use \eqref{cVl}. %\qed
\end{proof}

In the next Section we consider concrete examples for the operator
$L$.

\section{Examples}

As we have seen in the previous section, the sufficient conditions
\eqref{dVl}--\eqref{cVl} have identical structure for death, birth
and hopping parts. Therefore, to present explicit expressions for
$L^\tr_\ren$, $V^\tr$ and others we may proceed in the following
manner. Let $a (\ga)= (KA)(\ga)$, where $A$ is a measurable function
on $\Ga_0$; let $a_\eps=KA_\eps$ be some scaling of $a$ and $A$,
$\eps>0$. Below we consider different types of the function $a$
(linear, exponential etc.) and present possible scalings such that
for any $\{\eta,\xi\}\subset\Ga_0$
\begin{equation}
\exists\,\lim_{\varepsilon \rightarrow 0}\sum_{\omega \subset \eta
}\varepsilon ^{-\left\vert \xi \right\vert }A_\eps \left( \omega
\cup \xi \right) =\lim_{\varepsilon \rightarrow 0}\varepsilon
^{-\left\vert \xi \right\vert } A_\eps  \left( \xi \right)
=:A^{V}(\xi).\label{gencondA}
\end{equation}
And after that we may apply this results to the our situation when
$A_\eps$ depends additionally on $x,y\in\X$.

\begin{enumerate}
\item Let $a(\gamma)\equiv\alpha\in\R$. Then $A(\eta)=\alpha
\cdot0^\n$ and we don't need scaling at all: if $a_\eps=a$ then
$A_\eps(\eta)=\alpha \cdot 0^\n$ and \eqref{gencondA} holds with
$A^V(\xi)=\alpha \cdot 0^{|\xi|}$.

\item Let $a(\ga)=\sum\limits_{x\in\ga}f(x)$ with some $f:\X\mapsto\R$.
Then $A(\eta)=\chi_{\{\eta=\{x\}\}}f(x)$. We consider the scaling
$f\mapsto\eps f$ for which
\begin{align*}
&\lim_{\varepsilon \rightarrow 0}\sum_{\omega \subset \eta
}\varepsilon ^{-\left\vert \xi \right\vert }A_\eps \left( \omega
\cup \xi \right)=\lim_{\varepsilon \rightarrow 0}\sum_{\omega
\subset \eta }\varepsilon ^{-\left\vert \xi \right\vert }\chi
_{\left\{ \omega \cup \xi =\left\{
x\right\} \right\} } \eps f(x)  \\
=&\lim_{\varepsilon \rightarrow 0}\varepsilon ^{-\left\vert \xi
\right\vert }\chi _{\left\{ \xi =\left\{ x\right\} \right\} }\eps
f(x)+\lim_{\varepsilon \rightarrow 0}\sum_{x\in \eta }\eps f(x)
=\chi _{\left\{ \xi =\left\{ x\right\} \right\} }f(x)=:A^{V}(\xi).
\end{align*}

\item Let $a(\ga)=\exp{\bigl\{\sum\limits_{x\in\ga}f(x)\bigr\}}$, $f:\X\mapsto\R$.
Then $A(\eta)=e_\la(e^f-1,\eta)$. We consider the same scaling
$f\mapsto\eps f$ for which
\begin{align*}
&\lim_{\varepsilon \rightarrow 0}\sum_{\omega \subset \eta
}\varepsilon ^{-\left\vert \xi \right\vert }A_\eps\left( \omega \cup
\xi \right)  =\lim_{\varepsilon \rightarrow 0}\sum_{\omega \subset
\eta }\varepsilon ^{-\left\vert \xi \right\vert }e_{\lambda }\left(
e^{\eps f }-1,\omega \cup \xi \right)  \\
=&\lim_{\varepsilon \rightarrow 0}e_{\lambda }\left( \frac{e^{\eps
f}-1}{\eps},\xi \right)
\sum_{\omega \subset \eta }e_{\lambda }\left( e^{\eps f}-1,\omega \right)  \\
=&\lim_{\varepsilon \rightarrow 0}e_{\lambda }\left( \frac{%
e^{\eps f }-1}{\varepsilon },\xi \right) \lim_{\varepsilon
\rightarrow 0}e_{\lambda }\left( e^{\eps f},\eta \right)
=e_{\lambda }\left( f ,\xi \right) =:A^V(\xi).
\end{align*}

\item
Let $a(\ga)=\sum\limits_{x\in\ga}\sum\limits_{y\in\ga\setminus x}
g(x,y)$ for some (non-symmetric, in general) function $g$ on
$\X\times\X$.  Then
\begin{align*}
A\left( \eta \right) =&\sum_{x\in \eta } K_{0}^{-1}\Bigl(\sum_{y\in
\cdot }g (x,y) \Bigr) \left( \eta \setminus x\right)  =\sum_{x\in
\eta }\chi _{\left\{ \eta \setminus x=\left\{ y\right\}
\right\} }g\left( x,y\right)  \\
=&\chi _{\left\{ \eta =\left\{ x,y\right\} \right\} }[ g\left(
x,y\right) +g\left( y,x\right) ] .
\end{align*}
We consider the scaling $g\mapsto \eps^2 g$. Then
\begin{align*}
&\lim_{\varepsilon \rightarrow 0}\sum_{\omega \subset \eta
}\varepsilon ^{-\left\vert \xi \right\vert }A_\eps\left( \omega \cup
\xi \right)  \\=&\lim_{\varepsilon \rightarrow 0}\sum_{\omega
\subset \eta }\varepsilon ^{-\left\vert \xi \right\vert }\chi
_{\left\{ \omega \cup \xi =\left \{
x,y\right\} \right\} }\eps^2\left[ g\left( x,y\right) +%
g\left( y,x\right) \right]  \\
=&\lim_{\varepsilon \rightarrow 0}\varepsilon ^{-2}\chi _{\left\{
\xi =\left\{ x,y\right\} \right\} } \eps^2\left[ g\left( x,y\right) +%
g\left( y,x\right) \right]   \\
=&\chi _{\left\{ \xi =\left\{ x,y\right\} \right\} }\left[ g\left( x,y\right) +%
g\left( y,x\right) \right]=:A^V(\xi).
\end{align*}%

\item Let $a(\ga)=\sum\limits_{x\in\ga}f(x)\exp{\bigl\{\sum\limits_{y\in\ga\setminus x}g(x,y)\bigr\}}$, $f:\X\mapsto\R$, $g:\X\times\X\mapsto\X$.
Then
\[
A \left( \eta \right) =\sum_{x\in \eta }f\left( x\right) e_{\lambda
}\bigl( e^{g(x,\cdot) }-1,\eta \setminus x\bigr).
\]%
Let us consider the scaling $f\mapsto \eps f$, $g\mapsto \eps
g$. Then%
\begin{align*}
&\lim_{\varepsilon \rightarrow 0}\sum_{\omega \subset \eta
}\varepsilon ^{-\left\vert \xi \right\vert }A_\eps\left(
\omega \cup \xi \right)  \\
=&\lim_{\varepsilon \rightarrow 0}\sum_{\omega \subset \eta
}\varepsilon ^{-\left\vert \xi \right\vert }\sum_{x\in \omega \cup
\xi }\eps f(x)e_{\lambda }\left( e^{\eps g\left( x,\cdot
\right) }-1,\omega \cup \xi \setminus x\right)  \\
=&\lim_{\varepsilon \rightarrow 0} \sum_{\omega \subset \eta
}\sum_{x\in \omega } \eps f(x)e_{\lambda }\left( e^{\eps g\left(
x,\cdot \right) }-1,\omega \setminus x\right) e_{\lambda }\left(
\frac{e^{\eps g\left( x,\cdot \right)}-1}{\varepsilon },\xi \right)  \\
&+\lim_{\varepsilon \rightarrow 0}\sum_{\omega \subset \eta
}\sum_{x\in \xi }\varepsilon ^{-1}\eps f(x) e_{\lambda }\left(
e^{\eps g\left( x,\cdot \right) }-1,\omega \right) e_{\lambda
}\left( \frac{e^{\eps g\left( x,\cdot \right) }-1}{\varepsilon }%
,\xi \setminus y\right)  \\
=&\lim_{\varepsilon \rightarrow 0}\sum_{x\in \xi }f(x) e_{\lambda
}\left(
e^{\eps g\left( x,\cdot \right) },\eta \right) e_{\lambda }\left( \frac{%
e^{\eps g\left( x,\cdot \right) }-1}{\varepsilon },\xi \setminus
y\right)  \\
=&\sum_{x\in \xi }f\left( x\right) e_{\lambda }\left( g\left(
x,\cdot \right) ,\xi \setminus x\right) =:A^{V}(\xi).
\end{align*}%

\item Let $a(\ga)=\bigl(\sum\limits_{x\in\ga}f(x)\bigr)\exp{\bigl\{\sum\limits_{y\in\ga }g(y)\bigr\}}$, $f,g:\X\mapsto\R$.
Then
\begin{align*}
A\left( \eta \right)  =&\left(
\chi _{\left\{ \cdot =\left\{ x\right\} \right\} }f(x) \star e_{\lambda }\left( e^{g }-1,\cdot \right) \right) \left( \eta \right)  \\
=&\sum_{x\in \eta } f(x) e_{\lambda }\left( e^g-1,\eta
\right)+\sum_{x\in \eta } f(x) e_{\lambda }\left( e^{g }-1,\eta
\setminus x\right).
\end{align*}%
Let us consider the scaling $f\mapsto \eps f$, $g\mapsto \eps g$.
Then
\begin{align*}
&\lim_{\varepsilon \rightarrow 0}\sum_{\omega \subset \eta
}\varepsilon ^{-\left\vert \xi \right\vert }A_\eps\left(
\omega \cup \xi \right)  \\
=&\lim_{\varepsilon \rightarrow 0}\sum_{\omega \subset \eta
}\varepsilon ^{-\left\vert \xi \right\vert }\sum_{x\in \omega \cup
\xi } \eps f(x) e_{\lambda }\left(
e^{\eps g }-1,\omega \cup \xi \right)  \\
&+\lim_{\varepsilon \rightarrow 0}\sum_{\omega \subset \eta
}\varepsilon ^{-\left\vert \xi \right\vert }\sum_{x\in \omega \cup
\xi } \eps f(x) e_{\lambda }\left(
e^{\eps g}-1,\omega \cup \xi \setminus x\right)  \\
=&\lim_{\varepsilon \rightarrow 0}\sum_{\omega \subset \eta
}\sum_{x\in \omega } \eps f(x) e_{\lambda }\left( e^{\eps g
}-1,\omega \right) e_{\lambda
}\left( \frac{e^{\eps g }-1}{\varepsilon }%
,\xi \right)  \\
&+\lim_{\varepsilon \rightarrow 0}\sum_{\omega \subset \eta
}\sum_{x\in \xi }\eps f(x) e_{\lambda
}\left( e^{\eps g }-1,\omega \right) e_{\lambda }\left( \frac{%
e^{\eps g }-1}{\varepsilon },\xi \right)  \\
&+\lim_{\varepsilon \rightarrow 0}\sum_{\omega \subset \eta
}\sum_{x\in \omega } \eps f(x) e_{\lambda }\left( e^{\eps g
}-1,\omega \setminus x\right)
e_{\lambda }\left( \frac{e^{\eps g }-1}{%
\varepsilon },\xi \right)  \\
&+\lim_{\varepsilon \rightarrow 0}\sum_{\omega \subset \eta
}\sum_{x\in \xi }\varepsilon ^{-1}\eps f(x) e_{\lambda }\left(
e^{\eps g}-1,\omega \right) e_{\lambda
}\left( \frac{e^{\eps g}-1}{\varepsilon }%
,\xi \setminus x\right) \\
=&\sum_{x\in \xi }f(x) e_{\lambda }\left( g,\xi \setminus
x\right)=:A^{V}(\xi).
\end{align*}
\end{enumerate}

Now we consider different types of birth-and-death and hopping
models with rates which have one of the forms considered above.
Using explicit expressions for $A$ and scaling for each concrete
model we have the expression for $A_\eps$ (which is $D^{(\eps)} _x$,
$B^{(\eps)} _x$ or $C^{(\eps)} _{x,y}$) and may easily obtain
expressions for $\h{L}_\ren$ and $L^\tr_\ren$ from
\eqref{bad-renorm} or \eqref{hop-renorm}. Using expressions for
$A^V$ (which is $D^{V} _x$, $B^{V} _x$ or $C^{V} _{x,y}$) we may
obtain expression for $\h{V}$ and $V^\tr$ as well as the form of
$\upsilon$ also from the Propositions~\ref{undercond} and
\ref{undercond2}. Let us turn to these concrete examples. We present
the Vlasov-type equations only.
\begin{example}[Surgailis model]
This birth-and-death model describes independent appearing and
disappearing points from a configuration after exponentially
distributed random times. The corresponding dynamics was considered
in \cite{Sur1983}, \cite{Sur1984}; the generator may be given for
$F\in\Fc(\Ga)$
\[
(LF)(\ga)=m\sum_{x\in\ga}[F(\ga\setminus x)-F(\ga)]+\sigma\int_\X
[F(\ga\cup x)-F(\ga)]dx.
\]
The scaling $m\mapsto m$, $\sigma \mapsto \eps^{-1}\sigma$ leads us
to the following Vlasov-type linear equation
\[
\frac{\partial}{\partial t}\rho_t(x)=-m \rho_t (x)+\sigma.
\]
\end{example}

\begin{example}[Contact model]
This model was considered in \cite{KS2006} (for further
investigations see \cite{KKP2008}, \cite{FKK2009}). The model
describes independent death of the members of a configuration, and,
on the other hand, production of new members of the configuration by
the existing ones. This is the simplest model for ecological
population dynamics. Note that a similar model was considered
already in \cite{Dur1979} as a particular case of a spatial
branching process in continuum. The generator is given on $\Fc(\Ga)$
by the expression
\begin{align*}
(LF)(\ga)=&m\sum_{x\in\ga}[F(\ga\setminus x)-F(\ga)]\notag\\&
+\la\sum_{x\in\ga}\int_\X a(x-y) [F(\ga\cup y)-F(\ga)]dy.
\end{align*}
The described scaling $m\mapsto m$, $\lambda\mapsto \eps^{-1}
\lambda$, $a\mapsto \eps a$ (that means that $L=L_\eps$) provides
the linear Vlasov-type equation also
\[
\frac{\partial}{\partial t}\rho_t(x)=-m \rho_t (x)+\la (\rho_t\ast
a)(x).
\]
Here and below $\ast$ denotes the usual convolution in $\X$.
\end{example}

\begin{example}[Social model]
This model was considered in \cite{FK2009}. It describes
birth-and-death process with migration from some ``reservoir'' and
competition between members of a configuration. The generator is
given on $\Fc(\Ga)$ by the expression
\begin{align*}
(LF)(\ga)&=\sum_{x\in\ga}\sum_{y\in\ga\setminus
x}a(x-y)[F(\ga\setminus x)-F(\ga)]\notag\\&\quad +\sigma\int_\X
[F(\ga\cup x)-F(\ga)]dx.
\end{align*}
The described scaling $a\mapsto \eps a$, $\sigma\mapsto
\eps^{-1}\sigma$ provides the non-linear Vlasov-type equation:
\[
\frac{\partial}{\partial t}\rho_t(x)=- \rho_t (x)(\rho_t \ast
a)(x)+\sigma.
\]
\end{example}

\begin{example}[Bolker--Dieckmann--Law--Pacala model]\label{ex:BDLP}
This model of population ecology was considered in \cite{BP1997},
\cite{BP1999}, \cite{DL2000}. Rigorous mathematical studying of this
model was done in \cite{FKK2009}. The individual of a population may
die independently as well as due to competition for resources; any
individual may produce a new one also. The generator is given on
$\Fc(\Ga)$ by the expression
\begin{align*}
(LF)(\ga)&=\sum_{x\in\ga}\Bigl(m+\sum_{y\in\ga\setminus
x}a^-(x-y)\Bigr)[F(\ga\setminus x)-F(\ga)]\notag\\&\quad +\la
\sum_{x\in\ga}\int_\X a^{+}(x-y) [F(\ga\cup y)-F(\ga)]dy.
\end{align*}
The scaling $a^\pm\mapsto \eps a^\pm$, $m\mapsto m$, $\la\mapsto
\eps^{-1}\la$ gives the following non-linear Vlasov-type equation:
\[
\frac{\partial}{\partial t}\rho_t(x)=-m \rho_t (x)- \rho_t
(x)(\rho_t \ast a^{-})(x)+\la(\rho_t\ast a^{+})(x).
\]
Note that in the space-homogeneous case we obtain the logistic-type
equation
\[
\frac{d}{d t}\rho_t =\bigl(\la \langle a^{+}\rangle-m - \langle
a^{-}\rangle \rho_t \bigr) \rho_t,
\]
where $\langle a^{\pm}\rangle = \int_\X a^\pm (x) dx$. For a
rigorous proof of convergence in this scaling see \cite{FKK2010c}.
\end{example}

\begin{example}[Contact model with establishment]
In this model the above described contact dynamics is improved by
taking into account the depressive role of the establishment.
Namely, the probability for a newborn member to survive in a new
place is smaller if there are more particles near this new place.
 In the language of a generator we describe this by the following expression
\begin{align*}
(LF)(\ga)&=m\sum_{x\in\ga}[F(\ga\setminus x)-F(\ga)]\notag\\&\quad
+\la\sum_{x\in\ga}\int_\X a(x-y)e^{-\sum_{u\in\ga}\phi(y-u)}
[F(\ga\cup y)-F(\ga)]dy.
\end{align*}
The scaling  $m\mapsto m$, $\lambda\mapsto \eps^{-1} \lambda$,
$a\mapsto \eps a$, $\phi\mapsto\eps\phi$ provides the following
non-linear Vlasov-type equation
\[
\frac{\partial}{\partial t}\rho_t(x)=-m \rho_t
(x)+\la(a*\rho_t)(x)e^{-(\phi*\rho_t)(x)}.
\]
\end{example}

\begin{example}[Contact model with fecundity]
This model describes influence of competition for resources on birth
intensity. Namely, if there are many existing members near a
``parent'', the probability to sent offspring for it is smaller. We
consider the following expression for the generator
\begin{align*}
(LF)(\ga)&=m\sum_{x\in\ga}[F(\ga\setminus x)-F(\ga)]\notag\\&\quad
+\la\sum_{x\in\ga}e^{-\sum_{u\in\ga\setminus x}\phi(x-u)}\int_\X
a(x-y) [F(\ga\cup y)-F(\ga)]dy.
\end{align*}
The previous scaling $m\mapsto m$, $\la\mapsto \eps^{-1}\la$,
$a\mapsto \eps a$, $\phi\mapsto \eps\phi$ yields another non-linear
Vlasov-type equations
\[
\frac{\partial}{\partial t}\rho_t(x)=-m \rho_t
(x)+\la(a*(\rho_te^{-(\phi*\rho_t)}))(x).
\]
\end{example}

\begin{example}[Dieckmann--Law model] This model, as well as the model from Example~\ref{ex:BDLP}
describes ecological population evolution. However, appearing of new
offsprings is proportional to the number of existing members of a
population. The generator is given by the following expression
\begin{align*}
(LF)(\ga)&=\sum_{x\in\ga}\Bigl(m+\sum_{y\in\ga\setminus
x}a^-(x-y)\Bigr)[F(\ga\setminus x)-F(\ga)]\notag\\&\quad
+\sum_{x\in\ga}\int_\X a^{+}(x-y)\Bigl(\la+\sum_{u\in\ga\setminus
x}b(x-u)\Bigr) [F(\ga\cup y)-F(\ga)]dy.\notag
\end{align*}
Note that without competition ($a^-=0$) this model explodes, namely,
the mean value of the number of members in any bounded region
becomes infinite after finite time; otherwise, if the competition
kernel $a^-$ is ``stronger'' than the kernel $b$ this effect is
absent (for details see \cite{FK2010}). After scaling $a^\pm\mapsto
\eps a^\pm$, $m\mapsto m$, $b\mapsto \eps b$ and $1\mapsto
\eps^{-1}$ (before the whole birth term) we obtain the following
non-linear Vlasov-type equation
\[
\frac{\partial}{\partial t}\rho_t(x)=-m \rho_t (x)- \rho_t
(x)(\rho_t \ast a^{-})(x)+\la(\rho_t\ast a^{+})(x)+(((b\ast
\rho_t)\rho_t)\ast a^{+})(x).
\]
\end{example}

\begin{example}[Glauber $G^+$ dynamics] This model is a continuous analog
of the Glauber dynamics on a lattice. It was considered in a couple
of works, see, e.g., \cite{KL2005}, \cite{KLR2007}, \cite{KKZ2006},
\cite{KMZ2004}, \cite{FKKZ2010}, \cite{FKL2007}. The generator of
this model is given by
\begin{align*}
(LF)(\ga)&=\sum_{x\in\ga}[F(\ga\setminus x)-F(\ga)]\notag\\&\quad
+z\int_\X e^{-\sum_{u\in\ga}\phi(y-u)}[F(\ga\cup y)-F(\ga)]dy.
\end{align*}
Here $z>0$ is an activity parameter and $\phi$ is a pair potential.
This generator has a reversible measure, namely, the Gibbs measure
with parameters $z$ and $\phi$ (see, e.g., \cite{KL2005},
\cite{FKL2007} for details). The scaling $m\mapsto m$, $z\mapsto
\eps^{-1}z$, $\phi\mapsto \eps\phi$ yields the following non-linear
Vlasov-type equation
\[
\frac{\partial}{\partial t}\rho_t(x)=- \rho_t (x)+ze^{-(\rho_t \ast
\phi )(x)}.
\]
For a rigorous proof of the convergence in this scaling see
\cite{FKK2010b}.
\end{example}

\begin{example}[Glauber $G^-$ dynamics] This model is similar to the previous
one, see, e.g., \cite{KKM2008}, \cite{KLR2007}.
\begin{align*}
(LF)(\ga)&=\sum_{x\in\ga}e^{\sum_{u\in\ga}\phi(x-u)}[F(\ga\setminus
x)-F(\ga)]\notag\\&\quad +z\int_\X [F(\ga\cup y)-F(\ga)]dy.
\end{align*}
The same scaling as before yields the similar non-linear Vlasov-type
equation
\[
\frac{\partial}{\partial t}\rho_t(x)=- \rho_t (x)e^{(\rho_t \ast
\phi )(x)}+z.
\]
\end{example}

\begin{example}[Free Kawasaki] This simplest exactly solvable hopping model
was considered in \cite{KLR2008}. It describes independent jumps of
particles in the system. The generator is the following
\[
\left( LF\right) \left( \gamma \right) =\sum_{x\in \gamma }\int_{\mathbb{R}%
^{d}}a\left( x-y \right) \left[ F\left( \gamma \setminus x\cup
y\right) -F\left( \gamma \right) \right] dy.
\]
We do not need scaling at all to obtain the linear Vlasov-type
equation
\[
\frac{\partial }{\partial t}\rho _{t}\left( x\right) =\left( \rho
_{t}\ast a\right) \left( x\right) -\left\langle a\right\rangle \rho
_{t}\left( x\right)  .
\]
\end{example}

\begin{example}[Density dependent Kawasaki] In this model the intensity of a jump
is linearly proportional to the existing population. The generator
is given on $\Fc(\Ga)$ by the expression
\[
\left( LF\right) \left( \gamma \right) =\sum_{x\in \gamma }\int_{\mathbb{R}%
^{d}}a\left( x-y\right) \sum_{u\in \gamma }b\left( x,y,u\right)
\left[ F\left( \gamma \setminus x\cup y\right) -F\left( \gamma
\right) \right] dy.
\]
The scaling $b\mapsto \eps b$ provides the following non-linear
Vlasov-type equation
\begin{align*}
\frac{\partial }{\partial t}\rho _{t}\left( x\right)  =&\int_{\mathbb{R}%
^{d}}\rho _{t}\left( y\right) a\left( x-y\right)
\int_{\mathbb{R}^{d}}\rho
_{t}\left( u\right) b\left( y,x,u\right) dudy \notag\\
&-\rho _{t}\left( x\right) \int_{\mathbb{R}^{d}}a\left( x-y\right) \int_{%
\mathbb{R}^{d}}\rho _{t}\left( u\right) b\left( x,y,u\right) dudy.
\end{align*}
In particular, if $b(x,y,u)=b(x-u)$ then
\begin{equation*}
\frac{\partial }{\partial t}\rho _{t}\left( x\right)
=((\rho_t(\rho_t\ast b))\ast a )(x) - \langle a\rangle\rho
_{t}\left( x\right)(\rho_t\ast b)(x).
\end{equation*}
If $b(x,y,u)=b(y-u)$ then
\begin{equation*}
\frac{\partial }{\partial t}\rho _{t}\left( x\right)  =(\rho
_{t}\ast b)\left( x\right) (\rho _{t}\ast a)\left( x\right)-\rho
_{t}\left( x\right) (\rho_t\ast a\ast b)(x).
\end{equation*}
\end{example}

\begin{example}[Gibbs--Kawasaki] This hopping particles model was
considered, e.g., in \cite{KLR2007}. The generator is given by the
expression
\[
\left( LF\right) \left( \gamma \right) =\sum_{x\in \gamma }\int_{\mathbb{R}%
^{d}}a\left( x-y \right)e^{-E^\phi(y,\ga)} \left[ F\left( \gamma
\setminus x\cup y\right) -F\left( \gamma \right) \right] dy
\]
It has a family of reversible Gibbs measures with the potential
$\phi$ and any activity $z>0$. The scaling $\phi\mapsto\eps\phi$
gives the non-linear Vlasov-type equation of the form
\[
\frac{\partial }{\partial t}\rho _{t}\left( x\right)=\left( \rho
_{t}\ast a\right) \left( x\right) \exp \left\{ -\left( \rho_{t} \ast
\phi\right) \left( x\right) \right\} -\rho _{t}\left( x\right)
\left( a\ast \exp \left\{- \rho \ast \phi\right\} \right) \left(
x\right) .
\]
\end{example}
\begin{example}
In the last example we consider another type of dynamics. Let $L$
describe the generator of the diffusion dynamics (see, e.g.,
\cite{KLR2006}, \cite{KKR2004}), namely, for any smooth cylindrical
function
\[
(LF)(\ga) = \sum_{x\in\ga} \Delta_{x}F(\ga) -
\sum_{x\in\ga}\sum_{y\in\ga\setminus x} \left\langle
\nabla\phi(x-y),\nabla_x F\right\rangle,
\]
where $\Delta$ is a classical Laplace operator in $\X$ and $\nabla$
is a gradient in $\X$. Our approach covers this case also. It can be
shown that the scaling $\phi\mapsto\eps\phi$ provides the following
non-linear partial differential Vlasov-type equation
\begin{align*}
\frac{\partial }{\partial t}\rho _{t}\left( x\right)=
&\Delta\rho_t(x) -\int_{\X} \phi(x-y) \left\langle\nabla\rho_t(x),
\nabla\rho_t(y)\right\rangle dy\\&-\rho_{t}(x) \int_\X \left\langle
\nabla \phi(x-y), \nabla\rho_t (y) \right\rangle dy.
\end{align*}
\end{example}

\paragraph{Acknowledgement}
Yu.K. is very thankful to  Errico Presutti for several instructive
discussions of the scaling limit problems and his permanent friendly
support. It is a pleasure to express many thanks for fruitful
discussions to Carlo Boldrighini, Tobias Kuna, Joel Lebowitz, and
Sandro Pellegrinotti during the visit in Rome in October 2009.

The financial support of DFG through the SFB 701 (Bielefeld
University) and German-Ukrainian Project 436 UKR 113/94, 436 UKR
113/97 is gratefully acknowledged.

\end{document}